\documentclass[preprint,graybox,natbib,nosecnum,url,aastex,aas_macros,aastex62]{svmult}
\usepackage{graphicx}

\def\aj{{AJ}}                   % Astronomical Journal
\def\apj{{ApJ}}                 % Astrophysical Journal
\def\apjl{{ApJ}}                % Astrophysical Journal, Letters
\def\apjs{{ApJS}}               % Astrophysical Journal, Supplement
\def\ao{{Appl.~Opt.}}           % Applied Optics
\def\aap{{A\&A}}                % Astronomy and Astrophysics
\def\procspie{{Proc.~SPIE}}     % Proceedings of the SPIE

\begin{document}

\title*{On the Verge of an Astronomy CubeSat Revolution}
% Use \titlerunning{Short Title} for an abbreviated version of
% your contribution title if the original one is too long
\author{Evgenya L. Shkolnik}
% Use \authorrunning{Short Title} for an abbreviated version of
% your contribution title if the original one is too long
\institute{School of Earth and Space Exploration; Interplanetary Initiative -- Arizona State University, Tempe, AZ 85287. USA, \email{shkolnik@asu.edu}}

%
% Use the package "url.sty" to avoid
% problems with special charactersa
% used in your e-mail or web address
%
\maketitle

\abstract{ 
CubeSats are small satellites built in standard sizes and form factors, which have been growing in popularity but have thus far been largely ignored within the field of astronomy. 
When deployed as space-based telescopes, they enable science experiments not possible with existing or planned large space missions, filling several key gaps in astronomical research.  Unlike expensive and highly sought-after space telescopes like the Hubble Space Telescope (HST), whose time must be shared among many instruments and science programs, CubeSats can monitor sources for weeks or months at time, and at wavelengths not accessible from the ground 
such as the ultraviolet (UV), far-infrared (far-IR) and low-frequency radio. 
Science cases for CubeSats being developed now include a wide variety of astrophysical experiments, including exoplanets, stars, black holes and radio transients.  
Achieving high-impact astronomical research with CubeSats is becoming increasingly feasible with advances in technologies such as precision pointing, compact sensitive detectors, and the miniaturisation of propulsion systems if needed.  CubeSats may also pair with the large space- and ground-based telescopes to provide complementary data to better explain the physical processes observed. 
}

\section{A Disruptive \& Complementary Innovation}

Fifty years ago, in December 1968, National Aeronautics and Space Administration (NASA) put in orbit the first satellite for space observations, the Orbiting Astronomical Observatory 2. Since then, astronomical observation from space has always been the domain of big players. Space telescopes are usually designed, built, launched and managed by government space agencies such as NASA, the European Space Agency (ESA) and the Japan Aerospace Exploration Agency (JAXA).  Costs typically range from hundreds of millions to a few billion dollars (USD) and can take decades to complete.

New technologies can disrupt the status quo by challenging the current assumptions and opening up new avenues of research.  Small satellites (SmallSats) are spacecraft with masses less than 180 kg in a variety of shapes and sizes, which allow us to trade some capabilities, such as mirror size, for lower cost, shorter build times, and more frequent launch opportunities. CubeSats are a subset of SmallSats, and are built in standard units of 10 cm x 10 cm x 10 cm cubes, called a 1U, and typically weigh $<$2 kg. It is precisely this standardisation of size and subsequent CubeSat components, such as the spacecraft structure, electronics, power systems and communications modules, which allows them to be purchased commercially off-the-shelf, dramatically lowering build costs, and causing the recent spike in the CubeSat's popularity. This same drive towards miniaturisation and component standardisation has pushed technological development globally in the past decades, evident everywhere in our everyday lives. For example, small, embedded cameras started as expensive gadgets but, driven largely by smartphones, became progressively cheaper, even smaller and better-performing, enabling them in turn to be used more broadly, in everything from cars to drones and even CubeSats. This virtuous cycle of both commerce and innovation is now actively driving the CubeSat surge.

\begin{figure}
\centering
\includegraphics[trim=0cm 0cm 0cm 0cm,clip=false,width=0.8\textwidth]{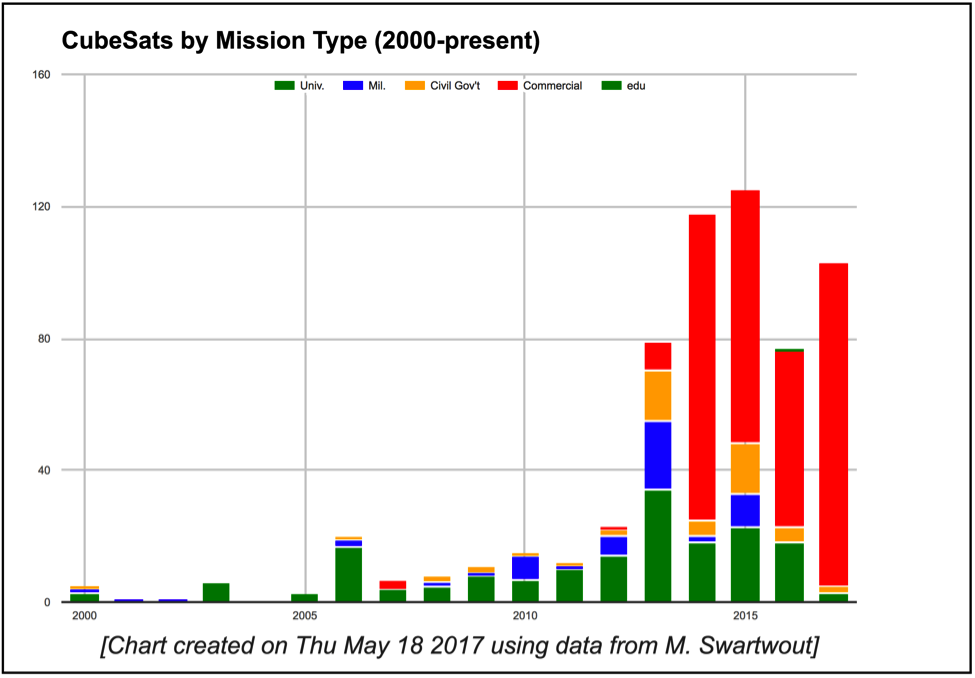}
\caption{A histogram counting the number of CubeSats launched each year. Data compiled up to May 18, 2017 and plot generated by by M. Swartwout of St. Louis University (https://sites.google.com/a/slu.edu/swartwout/home/cubesat-database) \label{histogram}
}
\end{figure}

CubeSats themselves benefit from this cycle. The CubeSat standard was proposed in 1999 by Jordi Puig-Suari of California Polytechnic State University and Bob Twiggs of Stanford University as an educational tool for teaching students about spacecraft hardware, electronics, and programming (e.g.~\cite{heid00}).  Universities built the majority of CubeSats before 2013, after which commercial usage overtook them, including much development done by the Aerospace Corporation \cite{hink09}.
The field is now dominated by tech startups whose culture for innovation and rapid improvement is well-suited to growing the CubeSat industry (Figure~\ref{histogram}).

Early CubeSats were custom affairs and limited in capability. As commercial demand increases, more capable components become available off-the-shelf, and the number of CubeSats in orbit continues to grow rapidly 
(https://www.isispace.nl/dutch-nanosatellite-company-gets-101-cubesats-launched-recordbreaking-pslv-launch/). 
The vast majority of launched CubeSats are used for telecommunications, technology demonstrations or navigation from low-Earth orbit (LEO), with only a few percent devoted to science, primarily for Earth observations (http://nanosats.eu/). 

CubeSats are launched as secondary payloads using a standard deployment system consisting of spring-loaded boxes, taking advantage of excess cargo space on launch vehicles. Being self-contained means the CubeSats do not add much risk, if any, to the primary payload. And as such, they fly for relatively little cost (e.g.~\cite{puig01}). In fact, the NASA's CubeSat Launch Initiative (CSLI\footnote{https://www.nasa.gov/directorates/heo/home/CubeSats\_initiative}) has been subsidising this cost for universities, high schools and non-profit organisations.

CubeSats will soon be traveling beyond Earth's orbit to explore the solar system. Increasing investment from space agencies and private industry is enabling planetary scientists to send specialized instruments housed in CubeSats to the Moon (e.g. LunaH-Map, \cite{west17}; Lunar Flashlight, \cite{cohe17}; Lunar IceCube, \cite{hein16}), Mars, (e.g. MarCo, \cite{kles16}), and even an asteroid (NEA-Scout, \cite{mcnu17}).  

The argument for CubeSats as a ``disruptive innovation'' for science is well laid out in Chapter 2 of a report produced by National Academies of Sciences, Engineering \& Medicine (NAS) entitled \textit{Achieving Science with CubeSats:
Thinking Inside the Box} \cite{nasreportcubesat}. The argument stems from the original definition by Clayton Christensen as the ``process by which a product or service takes root initially in simple applications at the bottom of a market and then relentlessly moves up-market, eventually displacing established competitors.''(https://hbr.org/product/disruptive-technologies-catching-the-wave-hbr-bestseller/95103-PDF-ENG).  Of course, CubeSats cannot displace the need for large space missions, such as the \$9.5B  James Webb Space Telescope (JWST) with its 6.5-m aperture and diverse suite of instruments. But there are many sufficiently bright sources for which smaller apertures, capturing spectral and temporal data, that can answer compelling scientific questions. 
For these, CubeSats can provide complementary approaches, such as time domain data, to the flagship observatories, thereby ushering in astronomy
to join the CubeSat boom.

%%%%%%%%%%%%%%%%%%%%%%%%%%%%%%%%%%%%%%%%%%%%%%%%%%%%%%

\section{Filling Science Gaps: Time Domain Astronomy Across the Electromagnetic Spectrum}

 All astrophysical phenomena change with time such that observed variability provides insight into the physical processes and motions at play. 
Many variable signals are best observed from space where conditions can be kept stable and where long and relatively unobstructed views are possible, compared to the Earth's day/night cycle. As recently evidenced by the remarkable stellar and exoplanet discoveries of NASA's Kepler mission\cite{thom17} (https://keplerscience.arc.nasa.gov/publications.html),  there is much to be learned from staring at one field for a long period of time, measured in weeks or months.   

Unlike ground-based time-domain surveys such as Pan-STARRS \cite{hube15} and the Large Synoptic Survey Telescope \cite{abel09}, space telescopes also allow access to energies across the electromagnetic spectrum inaccessible to these telescopes due to the absorption by the Earth's atmosphere, such as large gaps in the radio, the far-IR, and the entire high-energy range (UV to gamma rays). 
However, time domain programs with flagship space missions, such as HST and JWST, are challenging because the multi-purpose, multi-user, and multi-billion dollar telescopes are in such high demand.

\begin{figure}
\begin{center}
\includegraphics[trim=0cm 0cm 0cm 0cm,clip=false,width=0.8\textwidth]{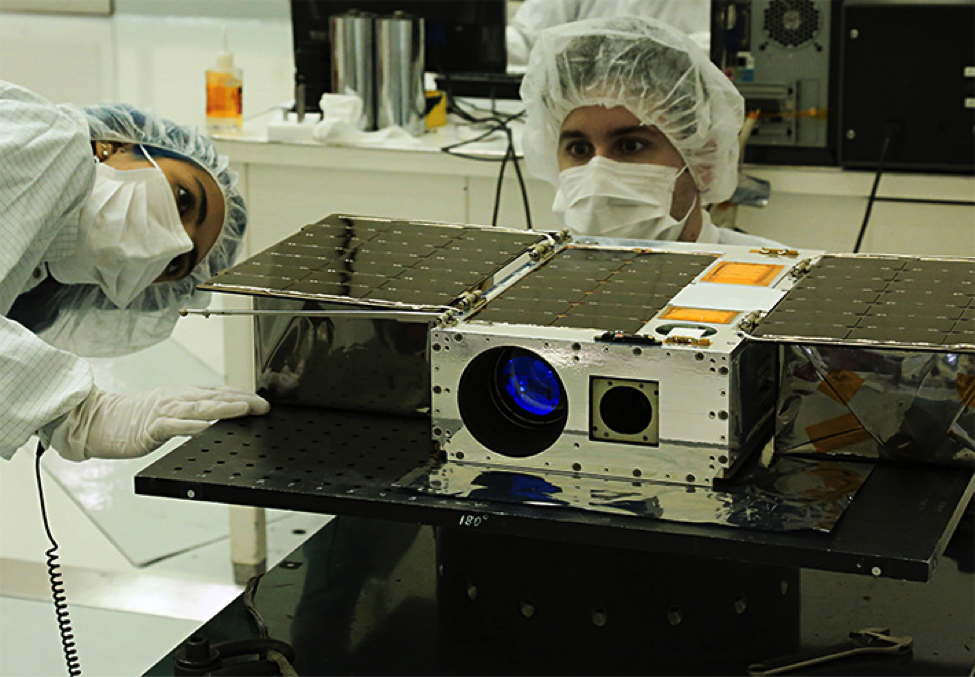}
\caption{Electrical Test Engineer Esha Murty (left) and Integration and Test Lead Cody Colley (right) prepare the ASTERIA spacecraft for mass properties measurements in April 2017 prior to spacecraft delivery. (Image credit: JPL)\label{asteria}
}
\end{center}
\end{figure}

There are now a handful of small observatories housed in CubeSats preparing for science:

$\bullet$ The ASTERIA (Arcsecond Space Telescope Enabling Research in Astrophysics) 6U CubeSat \cite{knap15}, led by the Jet Propulsion Laboratory (JPL) and the Massachusetts Institute of Technology (Principal Investigator (PI) Sara Seager), launched in August 2017 (Figure~\ref{asteria}).  ASTERIA's science goals are to measure exoplanetary transits across bright stars with $<$100 ppm photometry, making it the first CubeSat enabled for astronomical measurements. But more important are its technology goals to advance CubeSat capabilities for astronomy by achieving better than 5'' pointing stability over a 20-minute observation \cite{pong10}, and demonstrate milliKelvin-level temperature stability of the imaging detector. 

$\bullet$ PicSat, a French-led 3U CubeSat (PI Sylvestre Lacour) supported primarily by the European Research Council, was launched into a polar orbit in January 2018. Its primary goal is to observe in visible light the potential transit of the directly-imaged giant planet $\beta$ Pictoris b, and perhaps even its moons and debris (https://picsat.obspm.fr/).

$\bullet$    In 2016, NASA, through the Astrophysics Research and Analysis (APRA) program, funded its first astronomy CubeSat, HaloSat, a 6U CubeSat led by the University of Iowa (PI Philip Kaaret) \cite{kaar17}. HaloSat aims to measure the soft X-ray emission from the hot halo of the Milky Way galaxy to resolve the “missing baryon” problem, in which the number of baryons observed in the local universe is about half the amount recorded by the cosmic microwave background. These ``missing'' baryons may be residing in the hot halos around galaxies \cite{wang05,henl14}. 
HaloSat is expected to be launched on Orbital ATK mission AO-9 currently scheduled for May 2018 with support from the CSLI and NASA headquarters.

$\bullet$ In February of 2017, NASA funded its second astronomical 6U CubeSat, the Colorado Ultraviolet Transit Experiment (CUTE), led by the University of Colorado Boulder (PI Kevin France). It aims to conduct a survey of exoplanet transit spectroscopy in the near-UV \cite{flem17} of a dozen short-period, large planets orbiting FGK stars to constrain stellar variability and measure mass-loss rates. In previous UV transit spectroscopy observations carried out by HST, the stellar variability from transit to transit led to conflicting interpretations, possibly due to variations in stellar activity \cite{llam15,llam16}, and as such, many more transits are needed to disentangle the sources of variability.  CUTE is planning for a launch in the first half of 2020.

$\bullet$  This year, NASA funded two new astrophysics CubeSats. I am the PI of one, the Star-Planet Activity Research CubeSat (SPARCS), led by Arizona State University \cite{shko18}. It will be a 6U CubeSat devoted to the far- and near-UV monitoring of low-mass stars (0.2-0.6 M$_{\odot}$), the most dominant hosts of exoplanets \citep{dres15}.   The stellar UV radiation from M dwarfs is strong and highly variable \cite{mile17}, and impacts planetary atmospheric loss, composition and habitability \cite{rugh15,till17}. These effects are amplified by the extreme proximity of their habitable zones (HZ).  After a late-2021 launch to a sun-synchronous orbit, SPARCS will spend an entire month on each of at least a dozen M stars measuring rotational variability and flaring in both bands to be used as inputs to stellar atmosphere and planetary photochemistry models.

$\bullet$ BurstCube is the second one funded this year, led by the NASA's Goddard Space Flight Center (PI Jeremy Perkins), which also aims for a 2021 launch, to detect gamma ray transients in the 10-1000 keV energy range. Its fast reaction time and small localisation error are a valuable capability to catch the predicted counterparts of gravitational wave sources \cite{racu17}, complementing existing facilities such as Swift and Fermi. The team aims to eventually fly about 10 BurstCubes to provide all-sky coverage for significantly less cost than the typical large mission.

At this time, there are no funded far-IR CubeSats. Thermal stability and detector cooling needs to be considered at all wavelengths, but in the far-IR the detectors require extreme cooling to bring the thermal background to manageable levels. 
Cryocoolers capable of working within the current CubeSat power and space limitations have yet to be developed for astrophysics. IR Earth-observing CubeSats are currently leading the development in these regards, but with increased community interest towards far-IR astrophysics research with CubeSats \cite{farr17}, perhaps we will see more technological advances on the horizon. 

CubeSats like CUTE and SPARCS may also fill the upcoming lengthy gap in NASA's flagship UV capabilities. As HST's UV detectors degrade, there are no future opportunities planned until sometime after 2035, at least for the U.S. community, when the Large UV/Optical/IR Surveyor (LUVOIR) or the Habitable Exoplanet Imaging Mission (HabEx) may launch.

In addition to individual CubeSat science cases, pairing CubeSats - even those with a singular capability - with large telescopes can enhance their utility and impact. For instance, SPARCS may also be capable of `target-of-opportunity' UV observations for NASA's Transiting Exoplanet Survey Satellite (TESS) yield of rocky planets in M dwarf HZs. These will be some of the first HZ planets to be spectroscopically characterized by JWST and in need of the contemporaneous UV context for the interpretation of their transmission and emission spectra \cite{rugh15}. 

Even the large ground-based observatories can benefit from a CubeSat partner. Just two examples are electromagnetic follow-up conducted by CubeSats  to gravity wave detections \cite{evan17}, such as BurstCube plans, and the needed stellar activity monitoring by a dedicated CubeSat or two for the extreme-precision radial velocity searches for exoplanets by the upcoming extremely large telescopes \cite{gigu16}.  In each of these examples, the cost of the CubeSat is a small fraction of the total cost of the experiment, but it may provide the contextual data needed to best interpret the results.

%%%%%%%%%%%%%%%%%%%%%%%%%%%%%%%%%%%%%%%%%%%%%%%%%%%%%%

\section{Filling Technology Gaps}

Key technology developments can fill gaps \emph{for} CubeSats, as well as \emph{with} CubeSats for future missions.
Achieving compelling astronomy with CubeSats has recently become possible due to advances in  precision pointing, communications technology, deployables, and others summarised in Tables 5.1 and 5.2 of the NAS report \cite{nasreportcubesat}. Early CubeSats typically had short lifetimes once in orbit, lasting  only a few months.  With increased ground testing and added redundancies, the lifetimes have grown significantly. For instance,  AeroCube-4, built by the Aerospace Corporation (https://www.nasa.gov/sites/default/files/files/D\_Hinkley-Aerospace\_PICOSAT\_Capability\_Status\_2014.pdf), has been working in orbit for five years  and counting, with the rest of the AeroCube series (\#5 and greater) routinely flying for several years. 
In addition to the greater science potential, this increased reliability allows for CubeSats to be cost-effective space-borne testbeds for new technologies, lasting much longer and in more relevant radiation environments than sounding rockets (up to $\approx$30 minutes, $\approx$100 km) or balloons (up to $\approx$100 days, $\approx$ 50 km). 

One example, the Optical Communications and Sensor Demonstration (https://www.nasa.gov/directorates/spacetech/small\_spacecraft/ocsd\_project.html) aims to improve the range, accuracy and rate of communications using CubeSats, increasing downlink data rates from kb/s to Mb/s from LEO. For astronomy in particular, high-precision photometry requires high-precision pointing, which has been a great challenge for CubeSats. Four of the upcoming U.S.-led astronomy CubeSats discussed above are made possible by a commercially-available attitude control system (ACS), called XACT, specifically developed for CubeSats by Blue Canyon Technologies (BCT). 
This XACT ACS was first tested by the MinXSS solar CubeSat launched in December of 2015 \cite{maso16b}, which
demonstrated an improvement from tenths of degrees to a pointing RMS of 5'' to 15''. The is achieved with measurements taken by a suite of sensors including a star tracker, accelerometers, gyroscopes, a Sun sensor and a GPS receiver, which are combined to precisely drive three-axis reaction wheels, all of which fit into less than 1U of a 3U or 6U CubeSat. The first test of the XACT ACS on stars other than the Sun is being conducted by ASTERIA.

As the standard form factor of CubeSats gains popularity for astrophysical research, work on deployable and deformable mirrors in small packages is bolstered, benefiting both the science directly and  future space observatories that will use these components.
For example, Stiles et al. 2010 \cite{stil10} presented a concept in which a mirror unfolds from a 1U CubeSat volume to a 30 cm diameter primary mirror for a telescope, and Cahoy et al.'s \cite{caho12} CubeSat Deformable Mirror Demonstration (DeMi), shows how small deformable mirrors \cite{unde15} may become a key technology to correct optical system aberrations in high contrast. 

For these and many other innovations, CubeSats can play a critical role in increasing the Technology Readiness Levels (TRL; https://www.nasa.gov/directorates/heo/scan/engineering/technology/txt\_accordion1.html) of astronomy-enabling technologies. The TRL is a metric for assessing the maturity of a new technology. Increasing the TRL from one (beginning of basic research) to nine (proven to work in space) for a given technology requires teams to design, test, iterate, and validate. CubeSats excel here because of the quick turnaround time from idea to lessons learned at much lower costs.  

As an example, part of the SPARCS mission goals is to advance UV detectors by flying high quantum efficiency, UV-optimized detectors developed at JPL \cite{nikzad:2012}. The detectors on sounding rocket flights have already demonstrated greater than five times the sensitivity of those used by the most recent dedicated UV NASA mission, GALEX.  Flying these new detectors on a CubeSat will increase their TRL to the levels needed for future use in large space telescopes such as the next UV-enabled flagship mission.

A natural next step in the development of CubeSats for astronomy may be launching a constellation or swarm of CubeSats.  Such configurations will allow an increase in collecting area, or function as a single spectrometer where each CubeSat has its own photometric or spectroscopic wavelength range, or as a radio array of many CubeSat nodes \cite{lazi16,jone00,bana13,lucc17}. 
Already, five 20 cm x 20 cm x 20 cm, 7-kg cubes, albeit not actual CubeSats, fly together as part of the BRIght Target Explorer (BRITE) constellation \cite{hand16}. Each carries a 3-cm telescope with a wide field of view (about 24$^{\circ}$ on the sky) and has demonstrated high-precision (1.5 millimag) optical photometry of bright stars for asteroseismology \cite{baad16}. 
Beyond missions like this, swarms of CubeSats flying in formation become possible. University of Toronto's CanX-4 and 5 SmallSats, also both 20 cm x 20 cm x 20 cm cubes, successfully demonstrated formation flying after their launch in 2014 (https://directory.eoportal.org/web/eoportal/satellite-missions/c-missions/canx-4-5\#mission-status).  Formation flying of two CubeSats may be executed soon by the CANYVAL-X mission, short for the CubeSat Astronomy by NASA and Yonsei using Virtual Telescope Alignment eXperiment \cite{park17}.  CANYVAL-X aims to demonstrate their Vision Alignment System, which consists of an optic satellite to focus light from the sun to a separate detector satellite 10~m away.

%%%%%%%%%%%%%%%%%%%%%%%%%%%%%%%%%%%%%%%%%%%%%%%%%%%%%%%%%%%%%%%%%%%%%%%%%%%%%%
\section{On the Verge of Something Big with Something Small}

Currently, the cost of a research-grade astrophysics CubeSat is between \$5M and \$10M USD with a 2 -- 3 year start to launch time scale. 
Costs will continue to fall as more telescopes and instruments are developed to accommodate the CubeSat standard. The orders of magnitude difference in cost and development time of CubeSats compared to large missions opens the door to much broader participation. Smaller collaborations, individual institutions and countries seeking to get involved in space sciences may turn to CubeSats to grow their presence in space.  And, since astronomy is often a gateway science to STEM (science, technology, engineering and mathematics), CubeSat observatories may draw more people to these fields, providing opportunities for training students by giving them hands-on, end-to-end research and engineering experience in a reasonable time frame. 

As NASA, ESA and other global initiatives increase their solicitations for small astrophysics missions, technological and scientific advancement of CubeSats will pick up speed, and we will see a dramatic boost to CubeSat observatories, creating an even wider path towards the democratisation of space and adding a new dimension of how we do space-based astronomy.\\\\

%\begin{figure}
%\centering
%\includegraphics[trim=0cm 0cm 0cm 0cm,clip=false,width=0.8\textwidth]{atmos_transmission.png}
%\caption{The Earth's atmospheric windows include visible light and high-frequency radio waves. (Image credit:  TBD as we will create a better version of this with Nature Astronomy's graphics specialist.)\label{wavelength_gaps}
%}
%\end{figure}

%\begin{table*}
%\centering
%\includegraphics[trim=1.75cm 20cm 2cm 0cm,clip=true,width=1.\textwidth]{table.pdf}
%\caption{Table of potential challenges for CubeSat telescopes with current and developing solutions.}
%\label{fig:table}      
%\end{table*}

%% Put the bibliography here, most people will use BiBTeX in
%% which case the environment below should be replaced with
%% the \bibliography{} command.

% \begin{thebibliography}{1}
% \bibitem{dummy} Articles are restricted to 50 references, Letters
% to 30.
% \bibitem{dummyb} No compound references -- only one source per
% reference.
% \end{thebibliography}

%\bibliographystyle{naturemag}
%\bibliography{refs_master_2017Sept}

%% Here is the endmatter stuff: Supplementary Info, etc.
%% Use \item's to separate, default label is "Acknowledgements"

%\begin{addendum}
% \item[Correspondence] Correspondence and requests for materials
%should be addressed to E.L.S.~(email: shkolnik@asu.edu).
% \item E.L.S. has written this article in its entirety, but very much appreciates helpful comments from A. Dragushan, D. Jacobs and D. Ardila. 
% \item[Competing Interests] The author declares that she has no
%competing financial interests.
%\end{addendum}

%%
%% TABLES
%%
%% If there are any tables, put them here.
%%

\end{document}